\documentclass[twocolumn,showpacs,aps,prd]{revtex4}

\usepackage{epsfig}
\usepackage{graphics}
\usepackage{graphicx}
\usepackage{epic}
\usepackage{floatflt}
\usepackage{dcolumn}
\usepackage{amsmath}

\RequirePackage{xspace}

\usepackage{relsize}
\def\babar{\mbox{\slshape B\kern-0.1em{\smaller A}\kern-0.1em
    B\kern-0.1em{\smaller A\kern-0.2em R}}}

\def\Kbar  {\kern 0.2em\overline{\kern -0.2em K}{}\xspace}

\def\Kstarz  {\ensuremath{K^{*0}}\xspace}

\def\Kzb   {\ensuremath{\Kbar^0}\xspace}
\def\KzKzb {\ensuremath{K^0 \kern -0.16em \Kzb}\xspace}
\def\Dz    {\ensuremath{D^0}\xspace}

\def\Dbar  {\kern 0.2em\overline{\kern -0.2em D}{}\xspace}

\def\Dzb   {\ensuremath{\Dbar^0}\xspace}
\def\DzDzb {\ensuremath{D^0 {\kern -0.16em \Dzb}}\xspace}

\def\Bz    {\ensuremath{B^0}\xspace}
\def\B     {\ensuremath{B}\xspace}
\def\Bbar  {\kern 0.18em\overline{\kern -0.18em B}{}\xspace}

\def\Bzb   {\ensuremath{\Bbar^0}\xspace}
\def\Bu    {\ensuremath{B^+}\xspace}
\def\Bub   {\ensuremath{B^-}\xspace}

\def\BzBzb {\ensuremath{B^0 {\kern -0.16em \Bzb}}\xspace}
\def\BpBm    {\ensuremath{\Bu {\kern -0.16em \Bub}}\xspace}

\mathchardef\Upsilon="7107
\def\Y#1S{\ensuremath{\Upsilon{(#1S)}}\xspace}

\def\FourS {\Y4S}

\mathchardef\Deltares="7101
\mathchardef\Xi="7104
\mathchardef\Lambda="7103
\mathchardef\Sigma="7106
\mathchardef\Omega="710A
\def\Deltabar   {\kern 0.25em\overline{\kern -0.25em \Deltares}{}\xspace}
\def\Lbar {\kern 0.2em\overline{\kern -0.2em\Lambda\kern 0.05em}\kern-0.05em{}\xspace}
\def\Sigbar{\kern 0.2em\overline{\kern -0.2em \Sigma}{}\xspace}
\def\Xibar{\kern 0.2em\overline{\kern -0.2em \Xi}{}\xspace}
\def\Obar{\kern 0.2em\overline{\kern -0.2em \Omega}{}\xspace}
\def\Nbar{\kern 0.2em\overline{\kern -0.2em N}{}\xspace}
\def\Xb{\kern 0.2em\overline{\kern -0.2em X}{}}

\def\mes        {\mbox{$m_{\rm ES}$}\xspace}

\newcommand{\tev}{\ensuremath{\mathrm{\,Te\kern -0.1em V}}\xspace}
\newcommand{\gev}{\ensuremath{\mathrm{\,Ge\kern -0.1em V}}\xspace}
\newcommand{\mev}{\ensuremath{\mathrm{\,Me\kern -0.1em V}}\xspace}
\newcommand{\kev}{\ensuremath{\mathrm{\,ke\kern -0.1em V}}\xspace}
\newcommand{\ev}{\ensuremath{\mathrm{\,e\kern -0.1em V}}\xspace}
\newcommand{\gevc}{\ensuremath{{\mathrm{\,Ge\kern -0.1em V\!/}c}}\xspace}
\newcommand{\mevc}{\ensuremath{{\mathrm{\,Me\kern -0.1em V\!/}c}}\xspace}
\newcommand{\gevcc}{\ensuremath{{\mathrm{\,Ge\kern -0.1em V\!/}c^2}}\xspace}
\newcommand{\mevcc}{\ensuremath{{\mathrm{\,Me\kern -0.1em V\!/}c^2}}\xspace}

\def\mus  {\ensuremath{\rm \,\mus}\xspace}

\def\mus        {\ensuremath{\,\mu{\rm s}}\xspace}    
\def\gsim{{~\raise.15em\hbox{$>$}\kern-.85em
          \lower.35em\hbox{$\sim$}~}\xspace}
\def\lsim{{~\raise.15em\hbox{$<$}\kern-.85em
          \lower.35em\hbox{$\sim$}~}\xspace}

\def\CP                 {\ensuremath{C\!P}\xspace}

\def\to                 {\ensuremath{\rightarrow}\xspace}

\def\pep2{PEP-II}

\def\deltat{\ensuremath{{\rm \Delta}t}\xspace}

\xspace

\def\jetset74   {\mbox{\tt Jetset \hspace{-0.5em}7.\hspace{-0.2em}4}}

\def\bea{\begin{eqnarray}}
\def\eea{\end{eqnarray}}

\def\mes{\ensuremath{m_{\rm ES}}}
\def\de{\ensuremath{\Delta E}}

\def\bb{\ensuremath{{\sf B\overline{B}}}}

\def\fish    {\ensuremath{\cal F}\xspace}

\def\de {\ensuremath{\Delta E}}

\def\Kstarz {K^{*0}}

\def\bztdzksz {\ensuremath{B^0 \rightarrow {\Dtilde}^{0}K^{*0}}}

\def\bea{\begin{eqnarray}}
\def\eea{\end{eqnarray}}

\def\mes        {\mbox{$m_{\rm ES}$}\xspace}

\def\de{\ensuremath{\Delta E}}

\def\bb{{{B\bar{B}}}}

\def\Dtilde {\ensuremath{\tilde{D}}\xspace}

\def\bea{\begin{eqnarray}}
\def\eea{\end{eqnarray}}

\def\mes        {\mbox{$m_{\rm ES}$}\xspace}

\def\de{\ensuremath{\Delta E}}

\def\bb{{{B\overline{B}}}}

\def\Dtilde {\ensuremath{\tilde{D}}\xspace}

\newcommand{\BaBarType}      {PUB}  
\newcommand{\BaBarYear}       {08}
\newcommand{\BaBarNumber}     {037}
\newcommand{\SLACPubNumber} {13578}

\begin{document}
\begin{flushleft}
\babar-\BaBarType-\BaBarYear/\BaBarNumber \\ 
SLAC-PUB-\SLACPubNumber\\ 
\end{flushleft}
\title{Search for $b\to u$ transitions in $\Bz \to \Dz \Kstarz$ decays}
\author{B.~Aubert}
\author{M.~Bona}
\author{Y.~Karyotakis}
\author{J.~P.~Lees}
\author{V.~Poireau}
\author{E.~Prencipe}
\author{X.~Prudent}
\author{V.~Tisserand}
\affiliation{Laboratoire de Physique des Particules, IN2P3/CNRS et Universit\'e de Savoie, F-74941 Annecy-Le-Vieux, France }
\author{J.~Garra~Tico}
\author{E.~Grauges}
\affiliation{Universitat de Barcelona, Facultat de Fisica, Departament ECM, E-08028 Barcelona, Spain }
% ...
\collaboration{The \babar\ Collaboration}
\noaffiliation

\begin{abstract}\noindent
\pacs{13.25.Hw, 14.40.Nd}

We present a study of the decays $\Bz \rightarrow \Dz \Kstarz$ and $\Bz \rightarrow \Dzb \Kstarz$ 
with $\Kstarz \rightarrow K^{+}\pi^{-}$. 
The $\Dz$ and the $\Dzb$ mesons are reconstructed in the final states $f=K^+\pi^-$, 
$K^+\pi^-\pi^0$, $K^+\pi^-\pi^+\pi^-$ and their charge conjugates. 
Using a sample of $465$ million $B\overline{B}$ pairs collected with the \babar\ 
detector at the PEP-II asymmetric-energy  $e^+e^-$ collider at SLAC, 
we measure the ratio 
$R_{ADS}\equiv[\Gamma(\overline{B}^0\to[f]_D\overline{K}^{*0})+\Gamma(B^0\to[\bar{f}]_DK^{*0})]/[\Gamma(\overline{B}^0\to[\bar{f}]_D\overline{K}^{*0})+\Gamma(B^0\to[f]_DK^{*0})]$ for the three final states. 
We do not find significant evidence for a signal and set the following limits 
at 95$\%$ probability: $R_{ADS}(K\pi)<0.244$, $R_{ADS}(K\pi\pi^0)<0.181$ and 
$R_{ADS}(K\pi\pi\pi)<0.391$. 
From the combination of these three results, we find that the ratio $r_{S}$ 
between the $b \rightarrow u$ and the $b \rightarrow c$ amplitudes lies 
in the range $[0.07,0.41]$ at 95$\%$ probability.
\end{abstract}

\maketitle

Various methods have been proposed to determine the Unitarity Triangle angle 
$\gamma$~\cite{ref:GLW,ref:ADS,ref:DKDalitz} of the Cabibbo-Kobayashi-Maskawa (CKM) 
quark mixing matrix \cite{ref:ckm} using $\ensuremath{B^{-} \to \Dtilde^{(*)0}K^{(*)-}}$ 
decays, where the symbol $\Dtilde^{(*)0}$ indicates either a $D^{(*)0}$ or a 
$\Dbar^{(*)0}$ meson. A $B^-$ meson can decay into a 
$\Dtilde^{(*)0}K^{(*)-}$ final state via a $b\to c$ or a $b\to u$ process.  
\CP\ violation may occur due to interference between the amplitudes when 
the $D^{(*)0}$ and $\Dbar^{(*)0}$ decay to the same final state.  
These processes are thus sensitive to $\gamma=\mbox{arg}\{-V^{*}_{ub}V_{ud}/V^{*}_{cb}V_{cd}\}$. 
The sensitivity to $\gamma$ is proportional to the ratio between 
the $b \to u$ and $b \to c$ transition amplitudes ($r_B$), which depends on the $B$ decay channel 
and needs to be determined experimentally.  

\begin{figure}[htb]
\begin{center}
\epsfig{file=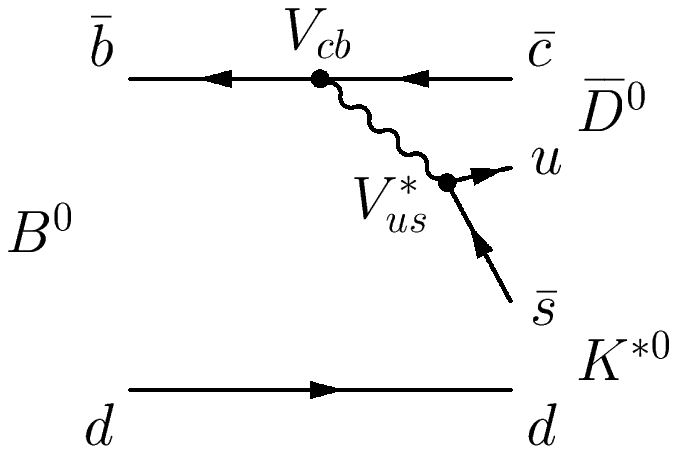,width=0.45\linewidth}
\epsfig{file=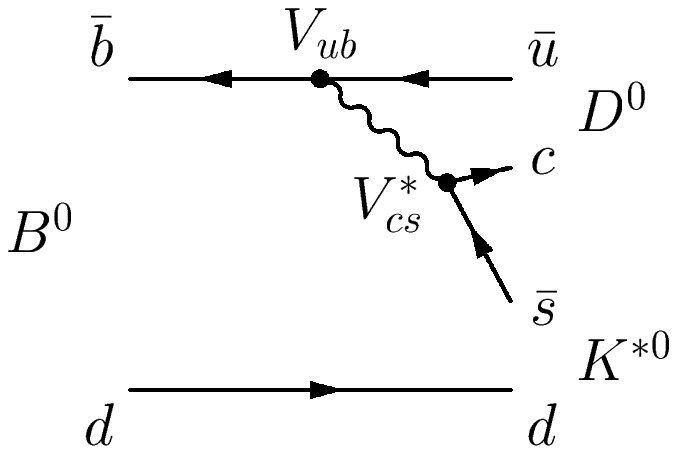,width=0.45\linewidth}
\end{center}
\caption{Feynman diagrams for $B^0 \rightarrow \Dzb \Kstarz$ (left, 
$\bar b \to \bar c$ transition) and $B^0 \rightarrow \Dz \Kstarz$ (right, $\bar b \to \bar u$ 
transition).\label{fig:feyn}}
\end{figure}

In this paper we consider an alternative approach, based on neutral $B$ mesons, 
which is similar to the ADS method \cite{ref:ADS} originally proposed for charged 
$B^{-} \to \Dtilde^{(*)0}K^{(*)-}$ decays. 
We consider the decay channel $\Bz \rightarrow \Dtilde^0 \Kstarz $ with 
$\Kstarz \to K^+ \pi^-$ (charge conjugate processes are assumed throughout the 
paper and $\Kstarz$ refers to the $K^{*}(892)^{0}$).  
This final state can be reached through $b \to c$ and $b \to u$ processes as shown 
in Fig.~\ref{fig:feyn}.  
The flavor of the $B$ meson is identified by the charge of the kaon 
produced in the $K^{*0}$ decay.
The neutral $D$ mesons are reconstructed in three final states, $f=K^+\pi^-$, 
$K^+\pi^-\pi^0$, $K^+\pi^-\pi^+\pi^-$. 
We search for $B^0 \to [\bar{f}]_D [K^+\pi^-]_{K^{*0}}$ events, where the CKM-favored 
$B^0 \to \Dzb K^{*0}$ decay, followed by the doubly Cabibbo-suppressed 
$\Dzb \to \bar{f}$ decay, interferes with the CKM-suppressed $B^0 \to \Dz K^{*0}$ 
decay, followed by the Cabibbo-favored $\Dz \to \bar{f}$ decay.  These are called 
``opposite-sign'' events because the two kaons in the final state have opposite 
charges. 
We also reconstruct a larger sample of ``same-sign'' events, which mainly 
arise from CKM-favored 
$\Bz \to \Dzb K^{*0}$ decays followed by Cabibbo-favored $\Dzb \to {f}$ decays.

In order to reduce the systematic uncertainties, we measure ratios of decay rates:
\begin{eqnarray}
R_{ADS} &\equiv& \frac{\Gamma(\overline{B}^0 \to [f]_D \overline{K}^{*0})
+\Gamma(B^0 \to [\bar{f}]_D K^{*0})}
{\Gamma(\overline{B}^0 \to [\bar{f}]_D \overline{K}^{*0})
+\Gamma(B^0 \to [f]_D K^{*0})} \label{eq:Radsbr}\\
{A}_{ADS} &\equiv&   \frac{\Gamma(\overline{B}^0 \to [f]_D\overline{K}^{*0})-\Gamma(B^0 \to [\bar{f}]_DK^{*0})}
{\Gamma(\overline{B}^0 \to [f]_D \overline{K}^{*0})+\Gamma(B^0 \to [\bar{f}]_D K^{*0})} 
\label{eq:Aadsbr}
\end{eqnarray}
where $R_{ADS}$ is the ratio between opposite- and same-sign events.  

The $K^{*0}$ resonance has a natural width (50 MeV/$c^2$) that is larger than the 
experimental resolution. This introduces a phase
difference between the various amplitudes.  We therefore 
introduce effective variables $r_S$, $k$, and $\delta_S$ \cite{gronau2002}, 
obtained by integrating over the region of the $B^0\to\Dtilde^0 K^+ \pi^-$ Dalitz plot 
dominated by the $\Kstarz$ resonance, defined as follows: 
\begin{eqnarray}
&& r_S^2\equiv\frac{\Gamma(\Bz \to \Dz K^+\pi^-)}{\Gamma(\Bz \to \Dzb K^+\pi^-)}
=\frac{\int dp\ A_{u}^2(p)}{\int dp\ A_{c}^2(p)},\label{eq:rs_square}\\
&& ke^{i\delta_S}\equiv\frac{\int dp\ A_{c}(p)A_{u}(p)e^{i\delta (p)}}{\sqrt{\int dp\ 
A_{c}^2(p) \ \int dp\ A_{u}^2(p)}}\,.\label{eq:k}
\end{eqnarray}
From their definition, $0\leq k \leq 1$ and $\delta_S\in[0,2\pi]$.
The amplitudes for the $b\to c$ and $b\to u$ transitions, $A_{c}(p)$ and $A_{u}(p)$, 
are real and positive and $\delta(p)$ is the relative strong phase. 
The variable $p$ indicates the position in the $\Dtilde^0 K^+ \pi^-$ Dalitz plot. 
The parameter $k$ accounts for contributions, in the $K^{*0}$ mass region, 
of higher-mass resonances. 
In the case of a two-body $\B$ decay, $r_S$ and $\delta_S$ become $r_B=A_{u}/A_{c}$ 
and $\delta_B$ (the strong phase difference between $A_{u}$ and $A_{c}$) with $k=1$.  
As shown in~\cite{violaDalitz}, the distribution of $k$ can be 
obtained by simulation studies based on realistic models for the 
different resonance contributions to the decays of 
neutral $B$ mesons into $\Dtilde^0K^{\mp}\pi^{\pm}$ final states.
When considering the region in the $B^0\to\Dtilde^0 K^+ \pi^-$ Dalitz plane where 
the invariant mass of the kaon and the pion is within 48\mevcc of the nominal 
$\Kstarz$ mass \cite{ref:PDG}, the distribution of $k$ is narrow, and is 
centered at 0.95 with a root-mean-square width of 0.03. 

Because of CKM factors and the fact that both diagrams in Fig.~\ref{fig:feyn} are color-suppressed, 
the average amplitude ratio $r_S$ in \bztdzksz\ is expected to be of order 0.3, 
larger than the analogous ratio 
for the charged $B^{-} \to D^{0(*)}K^{(*)-}$ decays, which is of order 
0.1~\cite{Bona:2005vz,ref:onlyforfairness}. 
This implies better sensitivity to $\gamma$ for the same number of events, 
an expectation that applies to all $B^{0} \to D^{0(*)}K^{(*)0}$ decays, 
and that motivates the use of neutral $B$ meson decays to determine $\gamma$. 
Currently, the experimental knowledge of $r_S$ \cite{ref:sha,violaDalitz} is $r_S<0.54$ at 95$\%$ probability.  

The ratios $R_{ADS}$ and $A_{ADS}$ are related to $r_S$, $\gamma$, $k$ and $\delta_S$ 
through the following relations:
\begin{eqnarray}
R_{ADS} &=& r_S^{2} + r_{D}^{2} + 2 k k_D r_S r_{D} \cos\gamma\cos(\delta_S + \delta_{D}), \label{eq:rads}\\ 
{A}_{ADS} &=& 2 k k_D r_S r_{D} \sin\gamma\sin(\delta_S + \delta_{D})/R_{ADS}\, ,
\label{eq:aads}
\end{eqnarray}
where
\begin{eqnarray}
&& r_D^2\equiv\frac{\Gamma(\Dz \to f)}{\Gamma(\Dz \to {\bar f})}
=\frac{\int dm \ A_{DCS}^2(m)}{\int dm\ A_{CF}^2(m)},\label{eq:rd_square}\\
&& k_D e^{i\delta_{D}}\equiv\frac{\int dm\ A_{CF}(m)A_{DCS}(m)e^{i\delta (m)}}{\sqrt{\int dm\ 
A_{CF}^2(m) \ \int dm\ A_{DCS}^2(m)}}\, ,\label{eq:kd}
\end{eqnarray}
with $0\leq k_D\leq 1$, $\delta_{D}\in[0,2\pi]$, $A_{CF}(m)$ and $A_{DCS}(m)$ 
the magnitudes of the Cabibbo-favored and the 
doubly-Cabibbo-suppressed amplitudes, $\delta(m)$ the relative 
strong phase, and the variable $m$ the position in the $D$ Dalitz plot. 
In the case of a two-body $D$ decay, $k_D=1$, $r_D$ is the ratio
between the doubly-Cabibbo-suppressed and the Cabibbo-favored decay
amplitudes and $\delta_D$ is the relative strong phase.

Determining $r_S$, $\gamma$ and $\delta_S$ from the measurements of $R_{ADS}$ 
and $A_{ADS}$, with the factor $k$ fixed, requires knowledge of the 
parameters ($k_D$, $r_{D}$, $\delta_D$), which depend on the specific neutral $D$ meson final 
states.  The ratios $r_D$ for the three $D$ decay modes 
have been measured \cite{ref:PDG}, as has the strong
phase $\delta_D$ for the $K\pi$ mode \cite{Asner:2008ft}. 
In addition, experimental information is available on $k_D$ and $\delta_D$ 
for the $K\pi\pi^0$ and $K\pi\pi\pi$ modes \cite{ref:CLEOc}.  
The smallness of the $r_D$ ratios 
implies good sensitivity to $r_S$ from a measurement of $R_{ADS}$. 
For the same reason, and since, with the present statistics, the asymmetries 
$A_{ADS}$ cannot be extracted from data, the sensitivity to $\gamma$ is 
reduced. 
The aim of this analysis is therefore the measurement of $r_S$. 
In the future, good knowledge of all the $r_D$, $k_D$ and $\delta_D$ 
parameters, and a precise measurement of the $R_{ADS}$ ratios 
for the three channels, will allow $\gamma$ and $\delta_S$ to 
be determined from this method as well.  

The results presented here are obtained with 423~fb$^{-1}$ of data collected at the $\FourS$ resonance 
with the \babar\ detector at the \pep2\ $e^+e^-$ collider at SLAC \cite{ref:pep2}, 
corresponding to 465 million $\bb$ events. 
An additional ``off-resonance'' data sample of 41.3~fb$^{-1}$, collected at a center-of-mass (CM) 
energy 40~\mev below the $\FourS$ resonance, is used to study backgrounds from continuum events, 
$e^+ e^- \to q \bar{q}$ ($q=u,d,s,$ or $c$).
The \babar\ detector is described elsewhere \cite{ref:det}.

The event selection is based on studies of off-resonance data and 
Monte Carlo (MC) simulations of continuum and $e^+e^-\to \Upsilon(4S)\to B \overline{B}$ events. 
All the selection criteria are optimised by maximising the 
function $S/\sqrt{S+B}$ on opposite-sign events, where $S$ and $B$ are the expected numbers 
of opposite-sign signal and background events, respectively.

The neutral $D$ mesons are reconstructed from a charged kaon and one or 
three charged pions and, in the $K\pi\pi^0$ mode, a neutral pion. 
The $\pi^0$ candidates are reconstructed from pairs of photon candidates, 
each with energy greater than 70\mev, 
total energy greater than 200\mev and invariant mass in the interval 
$[118,145]$\mevcc. The $\pi^0$ candidate's mass is subsequently constrained 
to its nominal value \cite{ref:PDG}. 

The invariant mass of the particles used to reconstruct the $D$ is required to 
lie within 14 MeV/$c^2$ ($\simeq$ 1.9$\sigma$), 20 MeV/$c^2$ ($\simeq$ 1.5$\sigma$) 
and 9 MeV/$c^2$ ($\simeq$ 1.6$\sigma$) of the nominal $D^0$ mass, 
for the $K\pi$, $K\pi\pi^0$ and $K\pi\pi\pi$ modes, respectively. 
For the $K\pi\pi\pi$ mode we also require that the tracks originate from 
a single vertex with a probability greater than 0.1\%.

The tracks used to reconstruct the $\Kstarz$ are constrained to originate 
from a common vertex and their invariant mass is required to 
lie within 48 \mevcc of the nominal $\Kstarz$ mass \cite{ref:PDG}. 
We define $\theta_{H}$ as the angle between the direction of flight of 
the $K$ and $B$ in the $\Kstarz$ rest frame. 
The distribution of $\cos\theta_{H}$ is proportional to $\cos^2\theta_{H}$ for
signal events and is expected to be flat for background events.
We require $|\cos\theta_{H}|>0.3$.  
The charged kaons used to reconstruct the $\Dtilde^0$ and $K^{*0}$ mesons are required to 
satisfy kaon identification criteria, based on Cherenkov angle and $dE/dx$ measurements 
and are typically 85\% efficient, depending on momentum and polar angle.  
Misidentification rates are at the 2\% level.

The $B^0$ candidates are reconstructed by combining a $\Dtilde^0$ and $K^{*0}$ candidate, 
constraining them to originate from a common vertex with a probability greater than
0.1\%. 
In forming the $B$, the $D$ mass is constrained to its 
nominal value \cite{ref:PDG}. 
The distribution of the cosine of the $B$ polar angle with respect to the
beam axis in the $e^+e^-$ CM frame $\cos\theta_{B}$ is expected to be
proportional to $1-\cos^2\theta_B$.  We require $|\cos\theta_{B}|<0.9$.
We measure two almost independent kinematic variables: the beam-energy substituted mass 
$\mes\equiv\sqrt{(E^{*2}_{0}/2+\vec{p_0}\cdot\vec{p_B})^2/E^{2}_{0}-{p_B}^2}$, and 
the energy difference $\de \equiv E^{*}_B-E^*_{0}/2$, where $E$ and $p$ are energy and momentum, 
the subscripts $B$ and $0$ refer to the candidate $B$ and $e^+e^-$ system, respectively, and the 
asterisk denotes the $e^+e^-$ CM frame.  
The distributions of $\mes$ and $\Delta E$ peak at the $B$ mass and zero, 
respectively, for correctly reconstructed $B$ mesons.  
The $B$ candidates are required to have 
$\Delta E$ in the range $[-16,16]$ MeV ($\simeq$ 1.3$\sigma$), $[-20,20]$ MeV 
($\simeq$ 1.5$\sigma$) and $[-19,19]$ MeV ($\simeq$ 1.4$\sigma$)
for the $K\pi$, $K\pi\pi^0$ and $K\pi\pi\pi$ modes, respectively.
Finally we consider events with \mes\ in the range $[5.20,5.29]$ \gevcc. 

We examine background $B$ decays that have the same final state 
reconstructed particles as the signal decay to identify modes 
with peaking structure in $\mes$ or $\Delta E$ that can potentially mimic 
signal events. 
We identify three such ``peaking background'' modes in the opposite-sign sample: 
$B^{0}\to D^{-}[K^{*0}K^{-}]\pi^{+}$ (for $K\pi$), 
$B^{0}\to D^{-}[K^{*0}K^{-}]\rho^{+}[\pi^{+}\pi^{0}]$ (for $K\pi\pi^0$) and 
$B^{0}\to D^{-}[K^{*0}K^{-}]a_1^{+}[\pi^{+}\pi^{+}\pi^{-}]$ (for $K\pi\pi\pi$). 
To reduce their contribution we veto all candidates for which the invariant mass of the 
$K^{*0}$ and the $K^{-}$ from the $D^0$ lies within $6$ \mevcc of the nominal $D^{-}$ mass.

After imposing the vetoes, the contributions of the peaking backgrounds 
to the $K \pi$, $K \pi \pi^0$ and $K\pi\pi\pi$ samples are predicted to be less than 
0.07, 0.05 and 0.12 events, respectively, at 95$\%$ probability. 
Other possible sources of peaking background are $B^{0}\to D^{0}\rho^{0}$ and 
$B^{0}\to D^{*-}[D^{0}\pi^{-}]\pi^{+}$, which contribute to the three decay 
modes in both the
same- and opposite-sign samples. These events could be reconstructed as signal, 
due to misidentification of a $\pi$ as a $K$. We impose additional restrictions 
on the identification criteria of charged kaons from $K^*$ decays to reduce the contribution of these
backgrounds to a negligible level.
Charmless $B$ decays, like $B^{0}\to K^{*0}K\pi$, can also contribute. 
The number of expected charmless background events, evaluated with data from the $\Dtilde^0$ mass
sidebands, is $N_{peak}$=0.5 $\pm$ 0.5 (0.1 $\pm$ 1.2) in the same (opposite) sign samples. 

In case of multiple $D$ candidate (less than $1\%$ of events), 
we choose the one with reconstructed $\Dtilde^0$ mass closest to the nominal mass \cite{ref:PDG}.
In the case of two $B$ candidates reconstructed from the
same $\Dtilde^0$, we choose the candidate with the largest
value of $|\cos\theta_{H}|$.

The overall reconstruction efficiencies for signal events are $(13.2 \pm 0.1)$\%, 
$(5.2 \pm 0.1)\%$ and $(6.5 \pm 0.1)\%$ for the $K \pi$, $K \pi \pi^0$ 
and $K \pi\pi\pi$ modes, respectively.

After applying the selection criteria described above, the remaining background is composed 
of continuum events and combinatorial $\bb$ events. 
To discriminate against the continuum background events (the dominant background component), 
which, in contrast to $\bb$ events, have a jet-like shape, we use a 
Fisher discriminant \fish\ \cite{ref:Fish}. The discriminant \fish\ is a linear combination 
of four variables calculated in the CM frame. 
The first discriminant variable is the cosine of the angle between 
the $B$ thrust axis and the thrust axis of the rest of the event. 
The second and third variables are $L_0=\sum_{i} p_i$, and $L_2 =\sum_{i} p_i  |\cos \theta_i|^2$, 
where the index $i$ runs over all the reconstructed tracks and energy deposits 
in the calorimeter not associated with a track, the tracks and energy deposits used to 
reconstruct the $B$ are excluded, $p_i$ is the momentum, and $\theta_i$ 
is the angle with respect to the thrust axis of the $B$ candidate. 
The fourth variable is $|\deltat|$, the absolute value of the measured proper time interval 
between the $B$ and $\bar{B}$ decays, calculated from the measured separation between 
the decay points of the $B$ and $\bar{B}$ along the beam direction.

The coefficients of \fish, chosen to maximize the separation between 
signal and continuum background, are determined using samples of simulated 
signal and continuum events and validated using off-resonance data. 

The signal and background yields are extracted, separately for each channel, by maximizing the extended likelihood 
$\mathcal{L} = (e^{- N'})/(N !)\cdot {N'}^{N}\cdot \prod_{j=1}^{N} f({\bf x}_{j} \mid {\bf \theta}, N')$.  
Here ${\bf x}_{j} =\{m_{ES}, \fish\}$, $\theta$ is a set of parameters, 
$N$ is the number of events in the selected sample and 
$N^\prime$ is the expectation value for the total number of events.   
The term $f({\bf x} \mid {\bf \theta}, N')$ is defined as :
\begin{eqnarray}
&&f({\bf x} \mid {\bf \theta}, N') N'= \frac{R_{ADS} N_{DK^*}}{1+ R_{ADS}} f_{SIG}^{OS}({\bf x}| {\bf \theta}_{SIG}^{OS}) \nonumber\\
&&+ \frac{N_{DK^*}}{1+ R_{ADS}} f_{SIG}^{SS}({\bf x}| {\bf \theta}_{SIG}^{SS}) +\nonumber\\
&&+  N_{bkg}^{OS}f^{OS}_{bkg}({\bf x}| {\bf \theta}^{OS}_{bkg}) + N_{bkg}^{SS} f^{SS}_{bkg}({\bf x}| {\bf \theta}^{SS}_{bkg})
\label{eq:pdf}
\end{eqnarray}
where $N_{DK^*}$ is the total number of signal events, 
$R_{ADS}$ is the ratio between opposite- and same-sign signal events, and 
``bkg'' refers to continuum or $B\overline{B}$ background, and 
$N_{cont}^{SS}$, $N_{cont}^{OS}$, $N_{B\overline{B}}^{SS}$, and $N_{B\overline{B}}^{OS}$ 
are the number of same- and opposite-sign events for continuum and $B\overline{B}$ 
backgrounds. 
The probability density functions (PDFs) $f$ are derived
from MC and are defined as the product of one-dimensional distributions of $m_{ES}$ 
and \fish. 
The $m_{ES}$ distributions are modeled with a Gaussian for signal, 
and threshold functions with different parameters for the 
continuum and $B\overline{B}$ backgrounds. 
The threshold function is expressed as follows:
\begin{eqnarray}
A(x) = x\,\sqrt{1 - (\frac{x}{x_{0}})^{2}}\cdot e^{c\,(1 - (\frac{x}{x_{0}})^{2})}\, ,
\end{eqnarray}
where $x_0$ represents the maximum allowed value for the variable $x$ 
described by $A(x)$ and $c$ accounts for the shape of the distribution. 
The \fish distributions are modeled with Gaussians. 

From the fit to data we extract $N_{DK^*}$, $R_{ADS}$, and the background yields 
($N_{cont}^{SS}$, $N_{cont}^{OS}$, $N_{B\overline{B}}^{SS}$, and $N_{B\overline{B}}^{OS}$). 
We allow the mean of the signal $m_{ES}$ PDF and parameters of the continuum 
$m_{ES}$ PDF's to float. 

The fitting procedure is validated using ensembles of simulated events. 
A large number of pseudo-experiments is generated with probability density 
functions and parameters as obtained from the fit to the data. 
The fitting procedure is then performed on these samples. 
We find no bias on the number of fitted events for any of the components. 

\begin{figure*}[htb]
\begin{center}
\epsfig{file=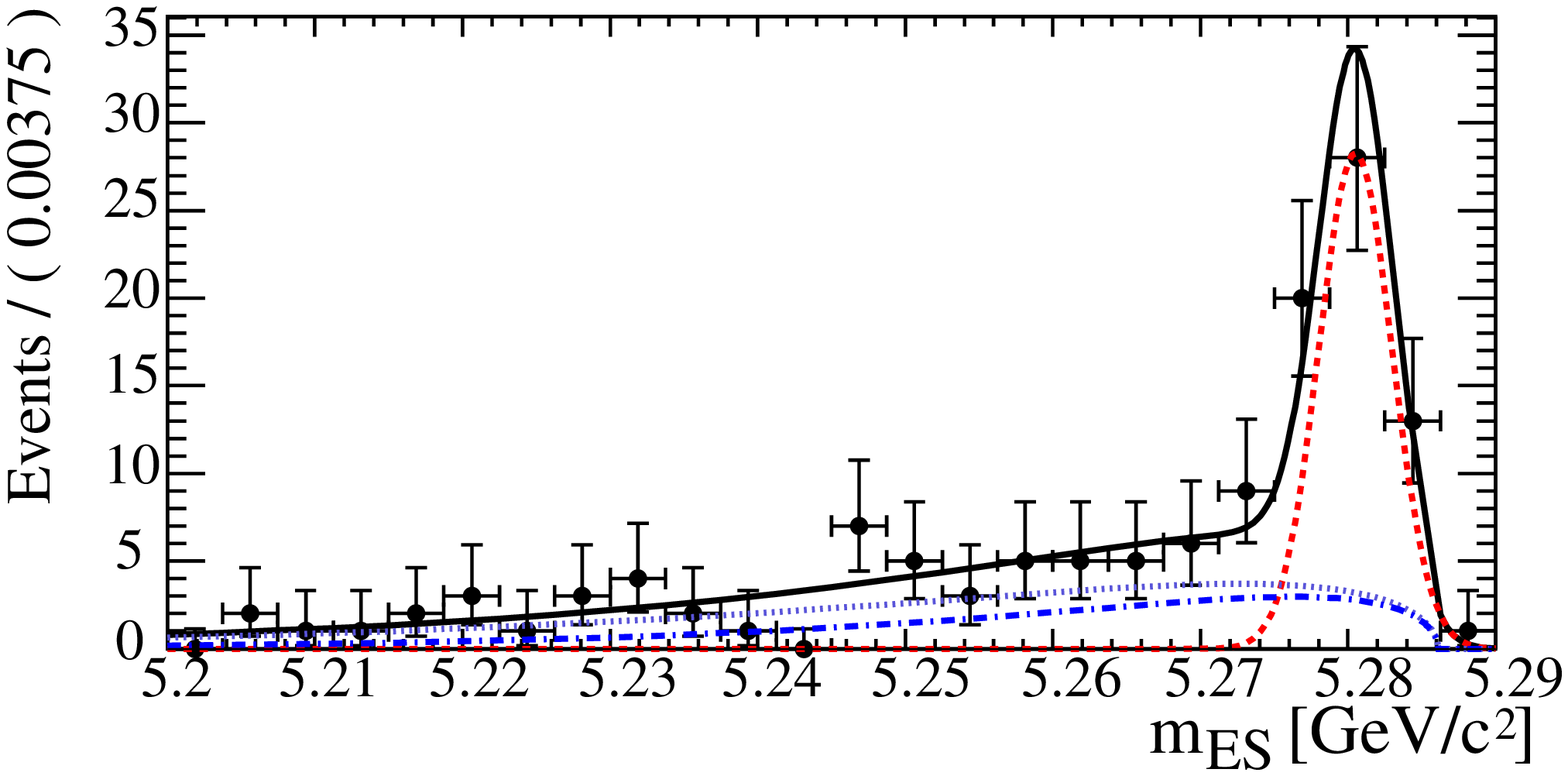,width=5.9cm}
\epsfig{file=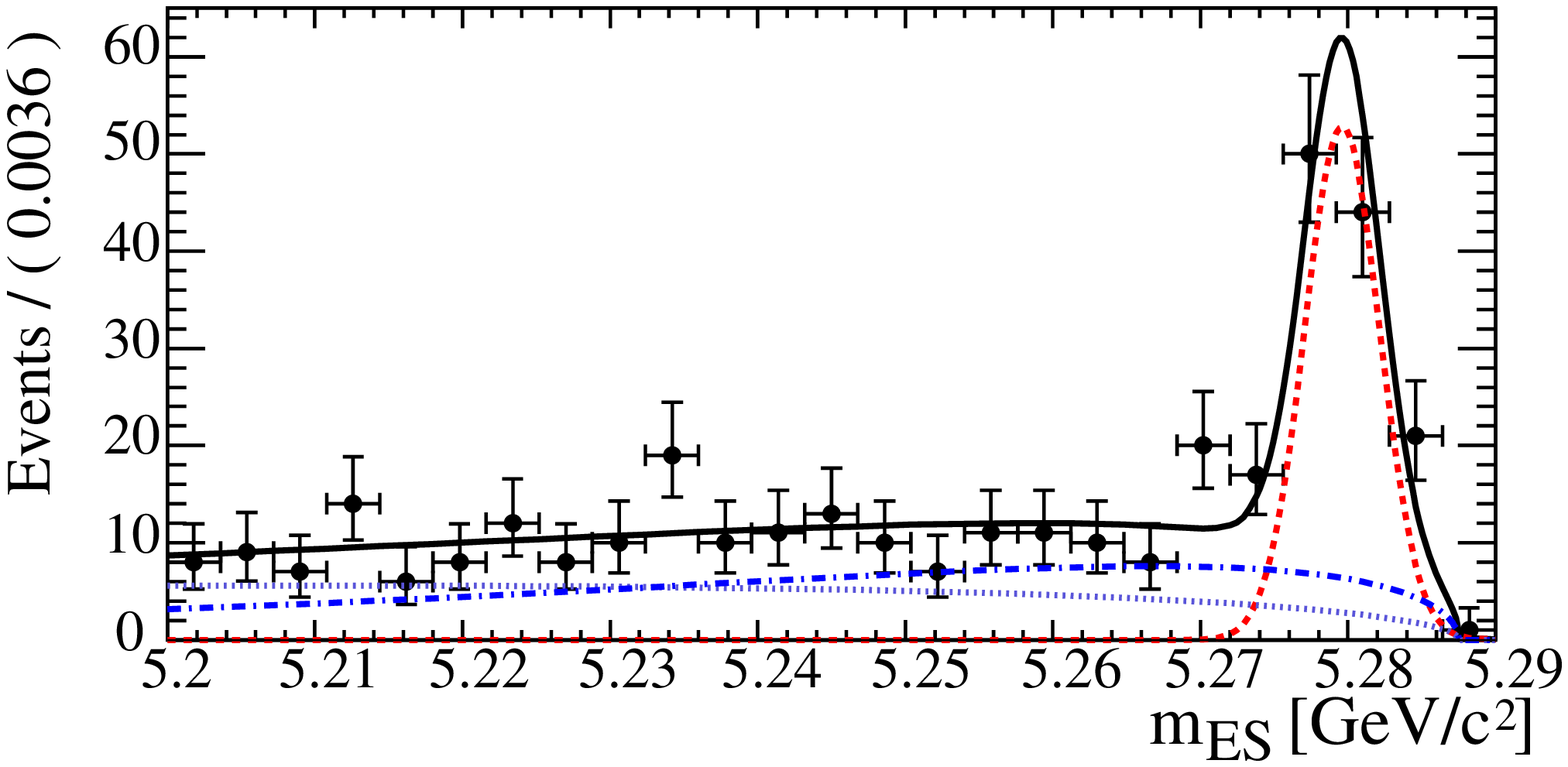,width=5.9cm}
\epsfig{file=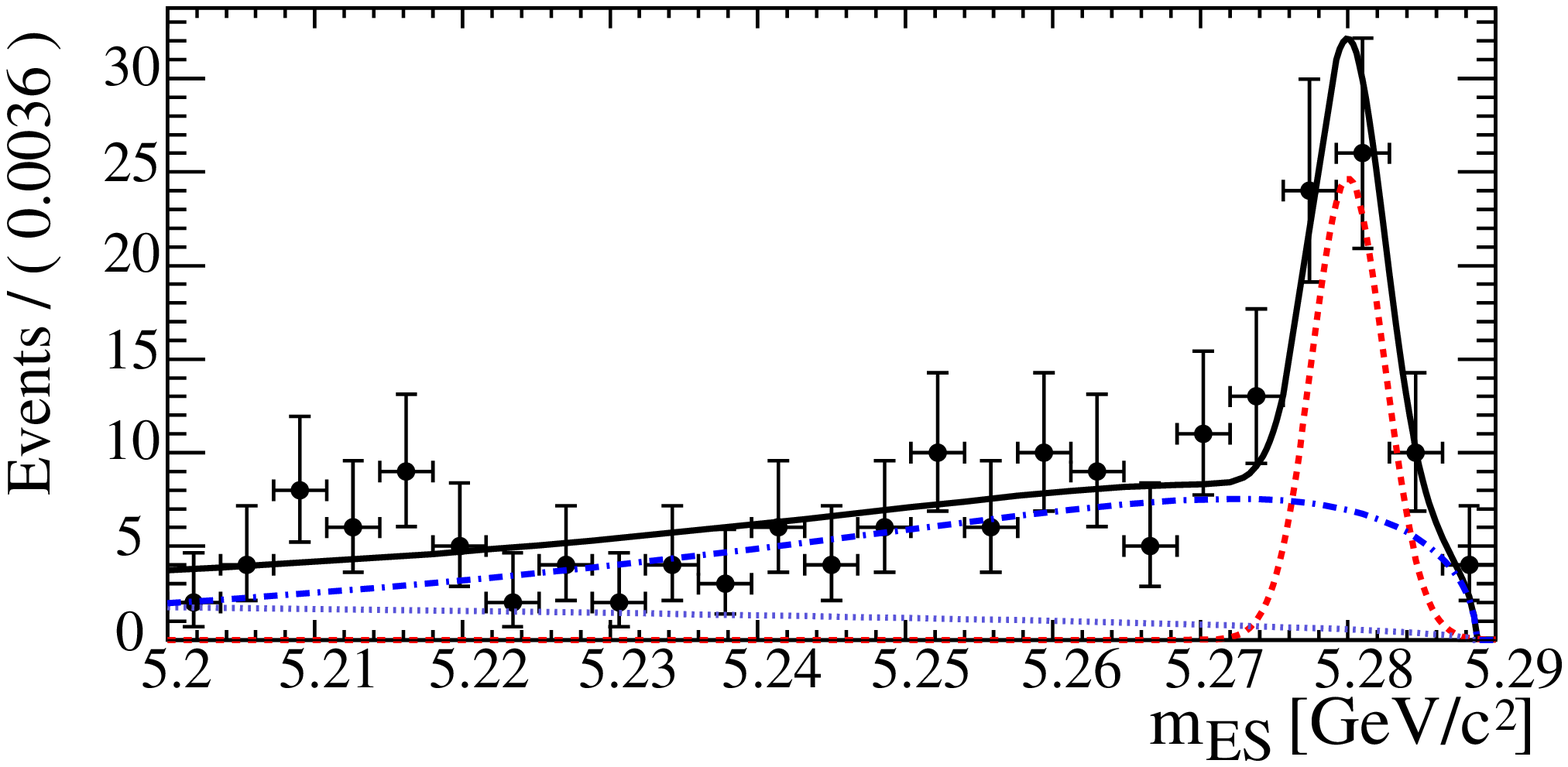,width=5.9cm}
\epsfig{file=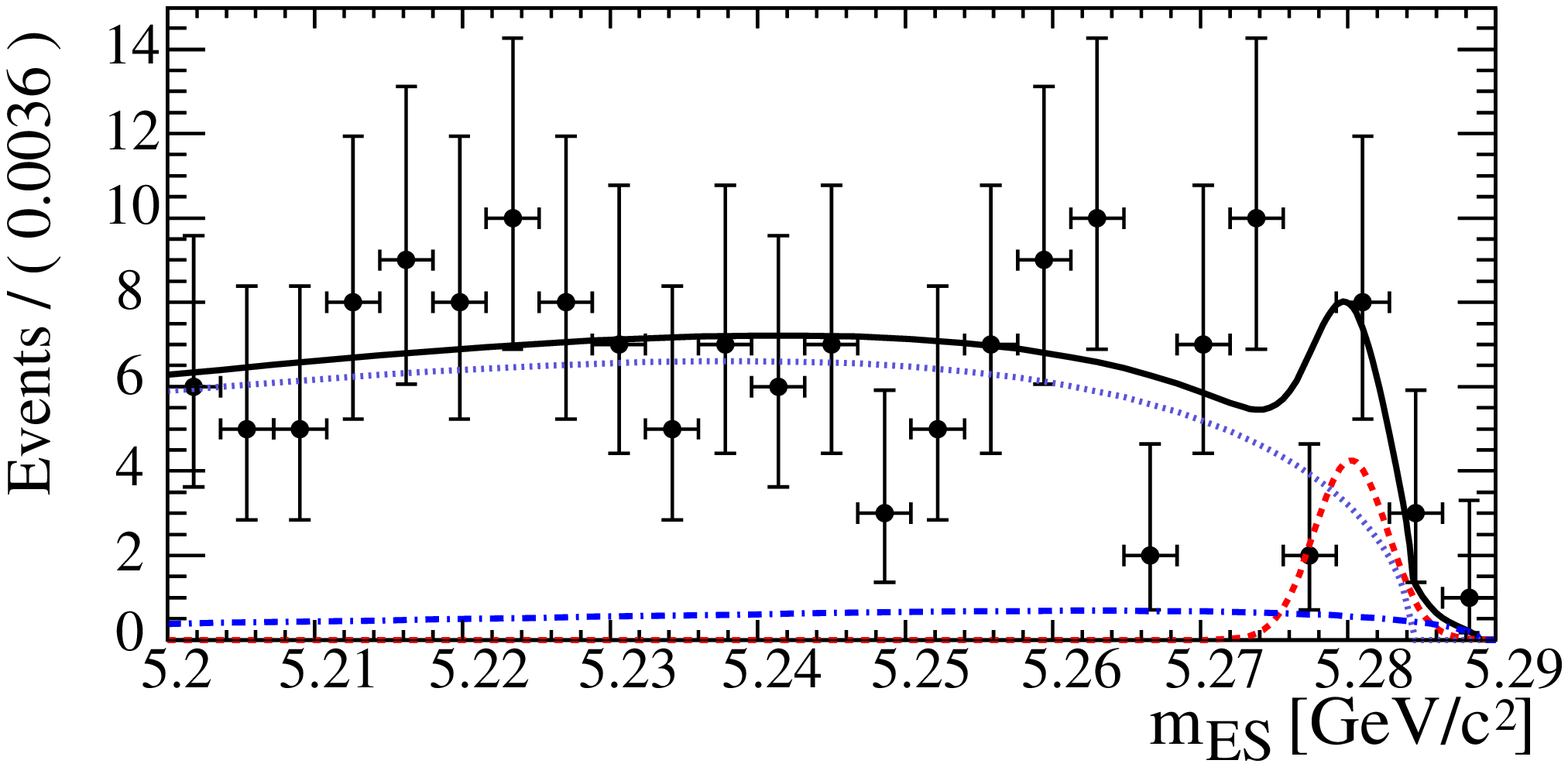,width=5.9cm}
\epsfig{file=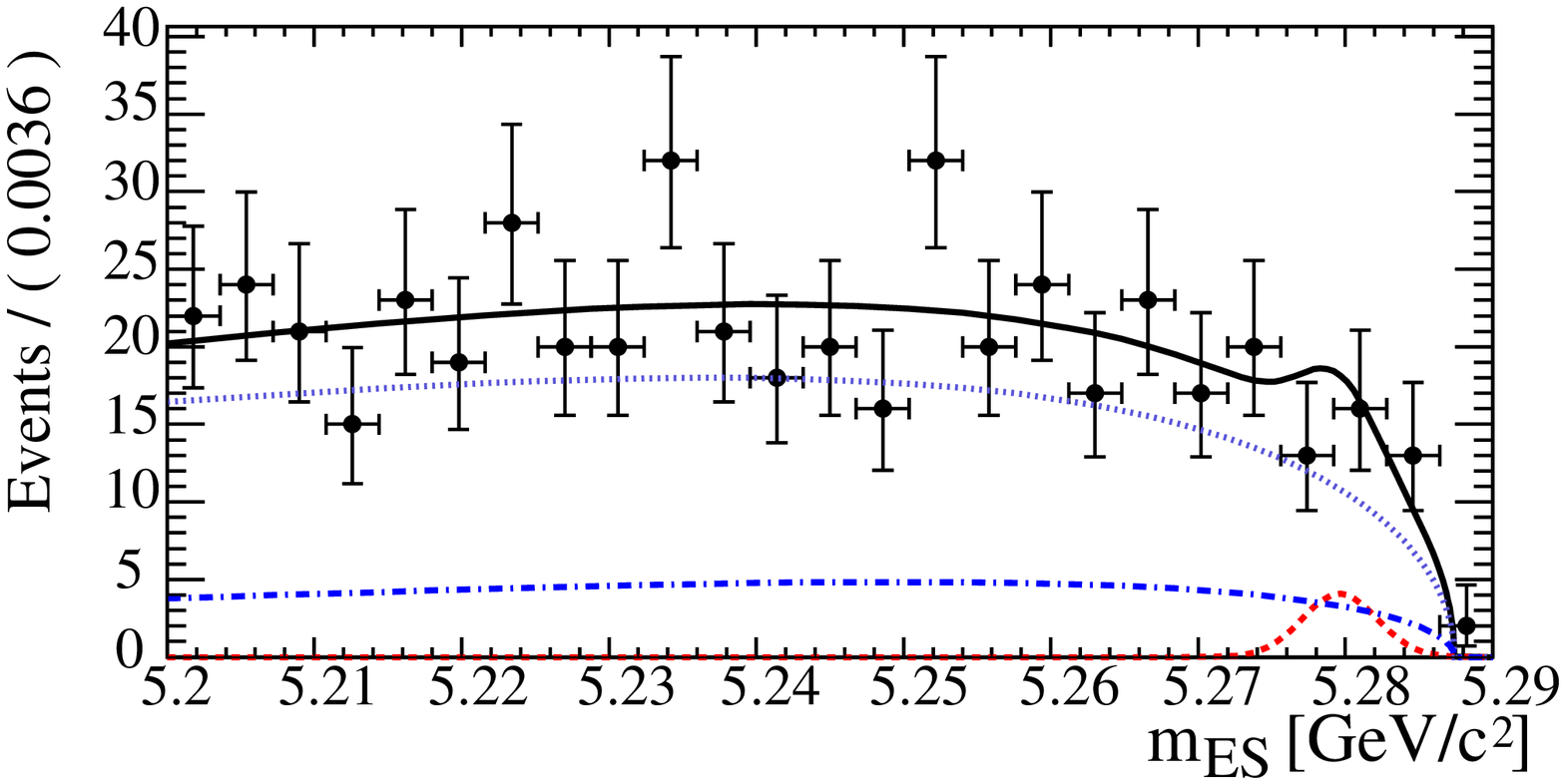,width=5.9cm}
\epsfig{file=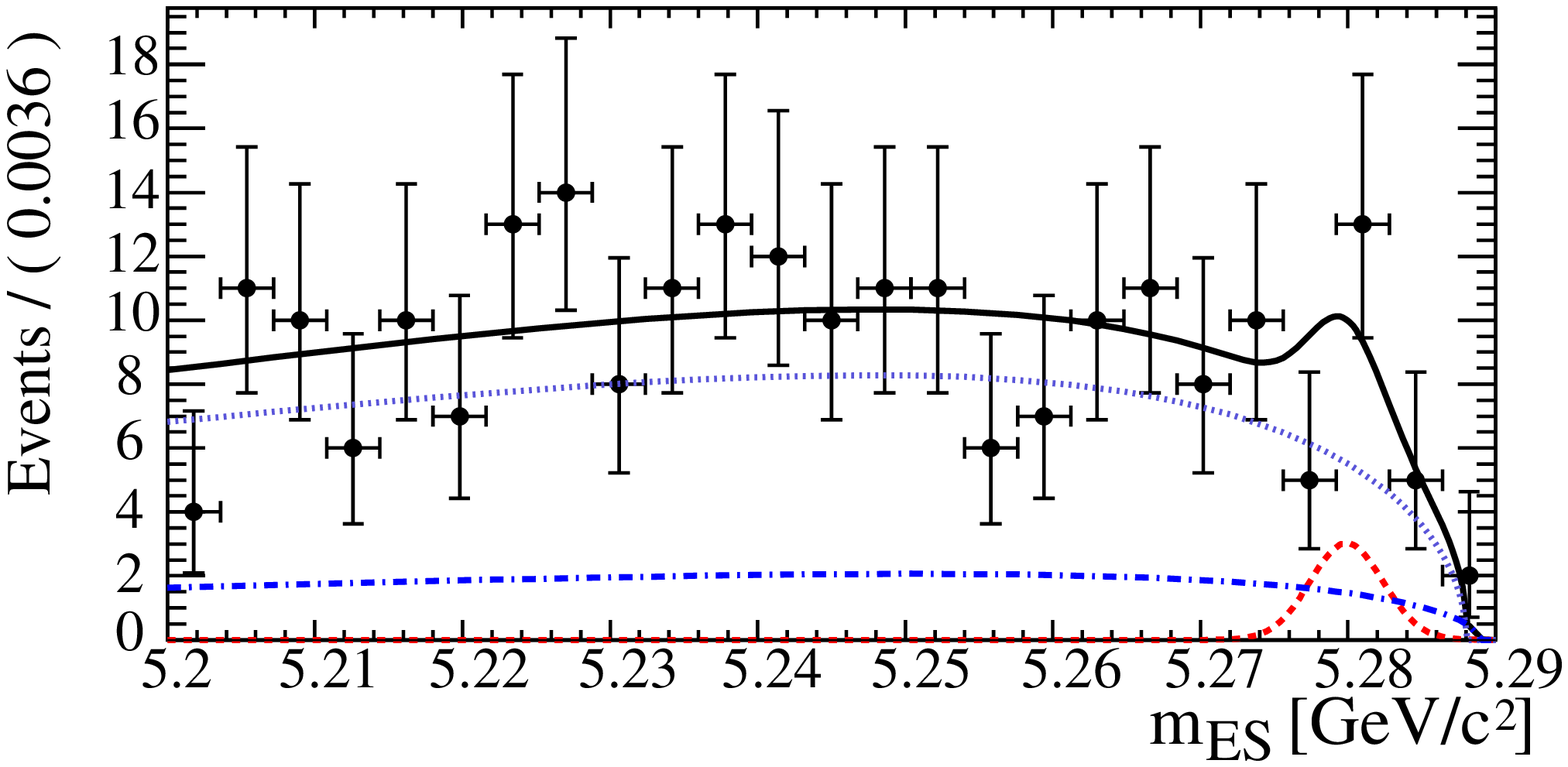,width=5.9cm}
\vspace{-0.4truecm}
\end{center}
\caption{ Projections of the fit onto the variable \mes after a cut on \fish is applied 
($>0.5$ for $K\pi$, $>0.7$ for $K\pi\pi^0$ and $>1$ for $K\pi\pi\pi$), 
to enhance the signal. The plots are shown for $K\pi$ (left), $K\pi\pi^0$ (middle) and 
$K\pi\pi\pi$ (right), same-sign (top) and opposite-sign (bottom) events. 
The points with error bars are data.
The dashed, dotted and dash-dotted lines represent the signal, continuum background and 
$\B \overline{B}$ background contributions, respectively. The solid line represents 
the sum of all the contributions. \label{fig:dataProj}}
\end{figure*}

The results for $N_{DK^*}$, $R_{ADS}$ and the background yields 
are summarized in Table~\ref{tab:fitres}. The total number of opposite-sign 
signal events in the three channels is $N_{SIG}^{OS}=24.4^{+13.7}_{-10.9}$ 
(statistical uncertainty only).
\begin{table}[htpb]
\caption{Fit results for $N_{DK^*}$, $R_{ADS}$ and the number of background events,
for the three channels. The uncertainties are statistical only. \label{tab:fitres}}
\begin{center}
\begin{tabular}{c c c c}
\hline\hline
channel  &    $K\pi$   &  $K\pi\pi^0$ &  $K\pi\pi\pi$ \\ \hline
$N_{DK^*}$  & $74\pm 12$  & $146\pm 17$ & $101\pm 17$     \\
$R_{ADS}$ & $0.067^{+0.070}_{-0.054}$  & $0.060^{+0.055}_{-0.037}$ & $0.137^{+0.113}_{-0.095}$ \\
$N_{B\overline{B}}^{SS}$ & $75\pm 16$    & $265\pm 33$  & $345\pm 35$   \\
$N_{B\overline{B}}^{OS}$  & $40\pm 17$    & $ 215\pm 41$ & $327\pm 48$   \\
$N_{cont}^{SS}$     & $387\pm 22$   & $2497\pm 56$ & $2058\pm 53$  \\
$N_{cont}^{OS}$      & $1602\pm 41$  & $7793\pm 96$ & $6372\pm 91$  \\\hline\hline
\end{tabular}
\end{center}
\end{table}
Projections of the fit onto the variable $m_{ES}$ are shown in Fig.~\ref{fig:dataProj} 
for the opposite- and same-sign samples.
To enhance the visibility of the signal, events are required to satisfy 
\fish$>0.5$ for $K\pi$, \fish$>0.7$ for $K\pi\pi^0$, and \fish$>1$ for $K\pi\pi\pi$. 
These requirements have an efficiency of about 67$\%$, 67$\%$ and 50$\%$ for signal 
and 9$\%$, 5$\%$ and 3$\%$ for continuum background. 

The systematic uncertainties on $R_{ADS}$ are summarized in Table \ref{tab:syst}. 
To evaluate the contributions related to the $m_{\rm ES}$ and \fish PDFs, 
we repeat the fit by varying all the PDF parameters that are fixed in the final fit 
within their statistical errors, as obtained from the parametrization on simulated 
events. 
To evaluate the uncertainty arising from the assumption of negligible 
peaking background contributions, we repeat the fit by varying the 
number of these events within their statistical errors. 
In this evaluation, we consider all the possible sources of such backgrounds, 
coming from charmless $B$ decays and from $B$ decays with a $D$ meson in 
the final state, as discussed above. 
For the multi-body $D$ decays, the selection efficiency on 
same- and opposite-sign events has been confirmed to be the same, 
regardless of the difference in the Dalitz structure, within a 
relative error of 3\%. 
Finally, a systematic uncertainty associated with cross feed between same- 
and opposite-sign events is evaluated from MC studies to be $(3.5 \pm 0.5)$\%, $(4.6 \pm 0.6)\%$ and $(1.9 \pm 0.4)\%$ 
for the $K \pi$, $K \pi \pi^0$ and $K \pi\pi\pi$ modes, respectively. 
The total systematic uncertainties are defined by adding the individual terms 
in quadrature. 
\begin{table}[htpb]
\caption{Systematic uncertainties $\Delta R_{ADS}$, in units of $[10^{-2}]$,
for $R_{ADS}^{K\pi}$, $R_{ADS}^{K\pi\pi^0}$ and $R_{ADS}^{K\pi\pi\pi}$. \label{tab:syst}}
\begin{center}
\begin{tabular}{c c c c}
\hline\hline
Source &    $K\pi$&  $K\pi\pi^0$&  $K\pi\pi\pi$\\ \hline
Sig. PDF  & $0.19$  & $0.11$ &$0.82$     \\
Cont. PDF & $0.32$  & $0.02$ &$0.29$    \\
$B\bar B$ PDF& $0.57$  & $0.16$ &$1.48$    \\
Peaking bkg & $1.70$  & $0.87$ &$1.40$   \\
$\epsilon_{CF}/\epsilon_{DCS}$& - & $0.17$& $0.39$ \\ \hline
cross-feed & $0.04$  & $0.05$ &$0.02$  \\
TOTAL                     & $1.8$ & $0.91$&$2.2$ \\\hline\hline
\end{tabular}
\end{center}
\end{table}
\begin{figure*}[htb]
\begin{center}
\vspace{-0.4truecm}
\epsfig{file=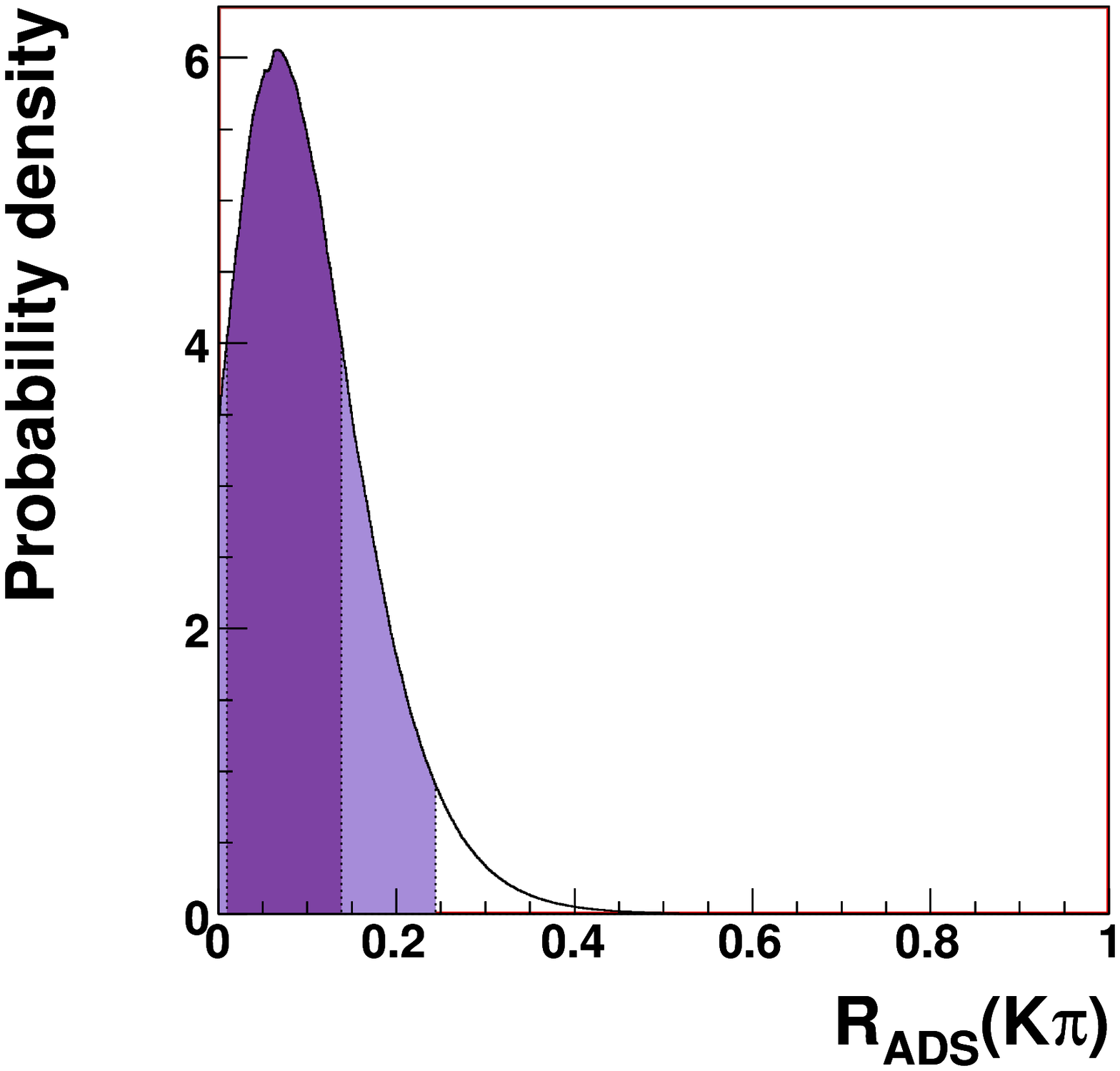,width=5cm}
\epsfig{file=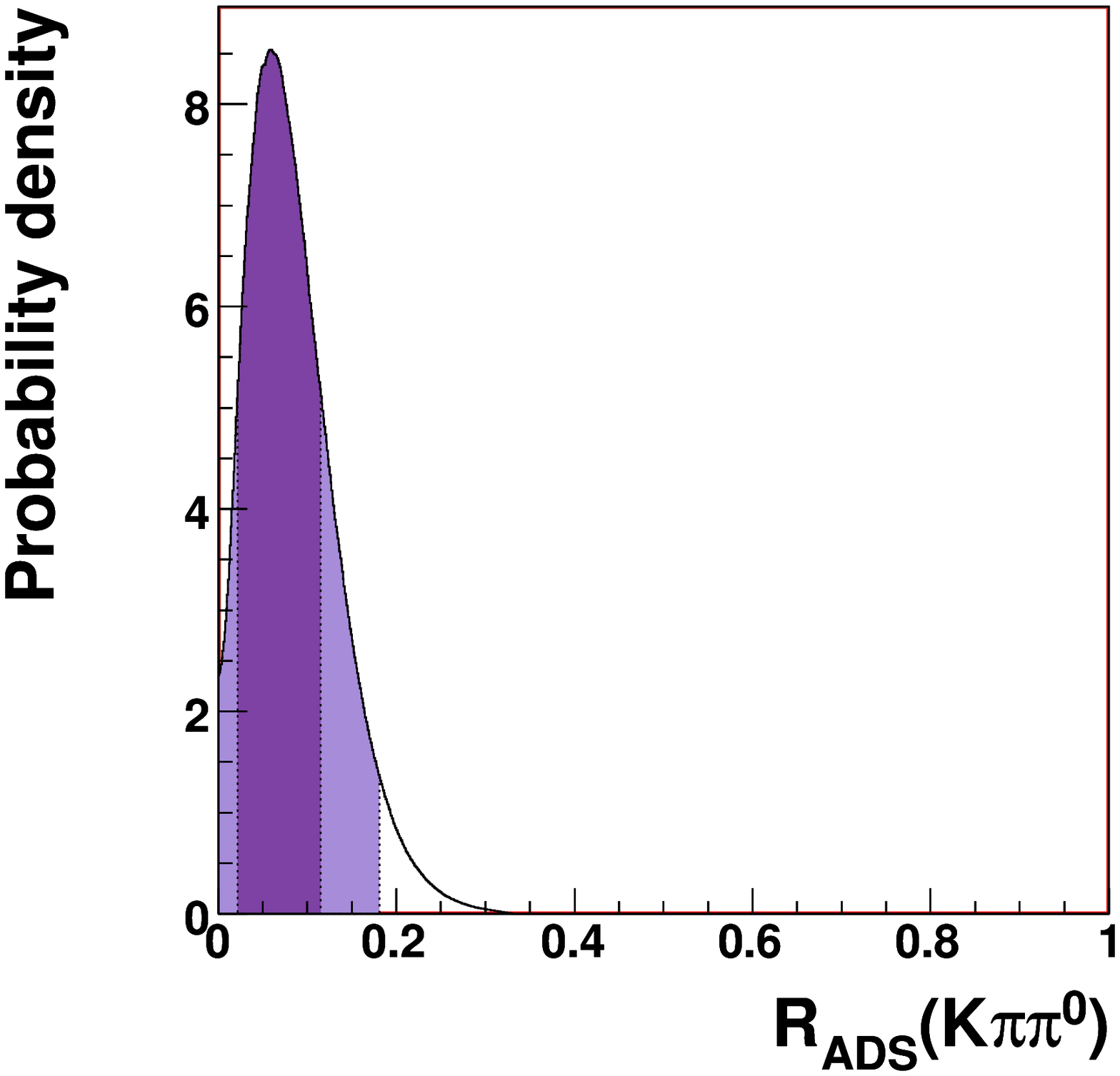,width=5cm}
\epsfig{file=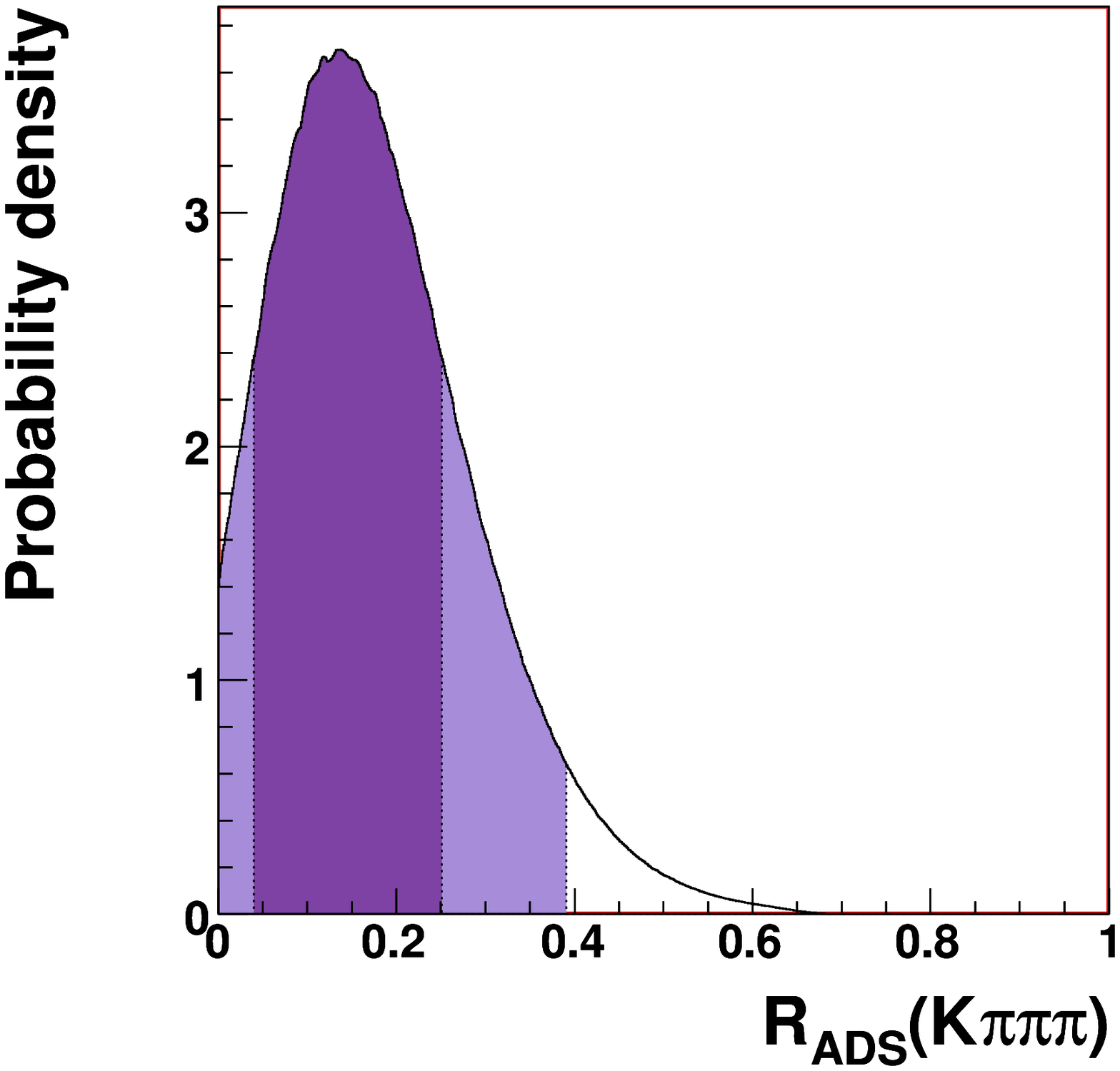,width=5cm}
\vspace{-0.4truecm}
\end{center}
\caption{Likelihood function for $R_{ADS}(K\pi)$ (left), 
$R_{ADS}(K\pi\pi^0)$ (middle) and $R_{ADS}(K\pi\pi\pi)$ 
(right), for $R_{ADS}\geq 0$, thus excluding unphysical values. 
The dark and light shaded zones represent the 68$\%$ and 95$\%$
probability regions, respectively. 
\label{fig:Rads_scan}}
\end{figure*}
\begin{figure}[h]
\begin{center}
\vspace{-0.4truecm}
\epsfig{file=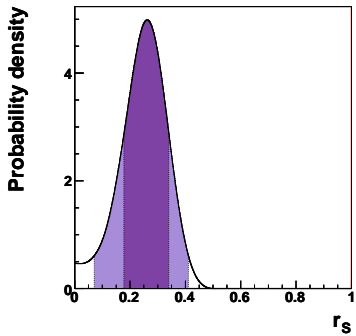,height=5.cm}
\vspace{-0.4truecm}
\end{center}
\caption{Likelihood function for $r_{S}$ from the combination of the measurements of 
$R_{ADS}$ obtained in the three $D$ decay channels. 
The dark and light shaded zones represent the 68$\%$ and 95$\%$ 
probability regions, respectively.
\label{fig:rS_threeCH}}
\end{figure}

The final likelihood ${\cal{L}}(R_{ADS})$ for each decay mode is obtained 
by convolving the likelihood returned by the fit with a Gaussian whose width equals 
the systematic uncertainty. 
Figure ~\ref{fig:Rads_scan} shows ${\cal{L}}(R_{ADS})$ for all 
three channels, where we exclude the unphysical region $R_{ADS}\leq 0$. 
The integral of the likelihood corresponding to $R_{ADS}<0$ is 9.5$\%$ for $K\pi$, 
15.8$\%$ for $K\pi\pi^0$ and 5.5$\%$ for $K\pi\pi\pi$. 
The significance of observing a signal is evaluated in each channel 
using the ratio 
$\log(\cal{L}_{\mbox{max}}/\cal{L}_{\mbox{0}})$, 
where $\cal{L}_{\mbox{max}}$ 
and 
$\cal{L}_{\mbox{0}}$ 
are the maximum likelihood values obtained from the 
nominal fit and from a fit in which the signal component is fixed to 
zero, respectively. 
We observe a ratio $R_{ADS}$ different from zero with a significance of 
1.1, 1.7 and 1.4 standard deviations for the $K\pi$, $K\pi\pi^0$ and 
$K\pi\pi\pi$ modes, respectively. 
Since the measurements for the $R_{ADS}$ ratios are not statistically significant, 
we calculate 95$\%$ probability limits by integrating the likelihoods, starting from $R_{ADS}=0$. 
We obtain $R_{ADS}(K\pi)<0.244$, $R_{ADS}(K\pi\pi^0)<0.181$ and 
$R_{ADS}(K\pi\pi\pi)<0.391$ at 95$\%$ probability. 
The overall significance of observing an $R_{ADS}$ signal, evaluated 
from the combination of the three measurements, is 2.5 standard deviations. 

Following a Bayesian approach, the measurements of the $R_{ADS}$ ratios 
are translated into a likelihood for $r_S$. 
A large number of simulated experiments for the parameters on which $R_{ADS}$ depends 
(see Eq.~\ref{eq:rads}) are performed.  For each experiment, the values of 
$R_{ADS}(K\pi)$, $R_{ADS}(K\pi\pi^0)$ and $R_{ADS}(K\pi\pi\pi)$ are obtained and a weight 
${\cal{L}}(R_{ADS}(K\pi)) {\cal{L}}(R_{ADS}(K\pi\pi^0))  {\cal{L}}(R_{ADS}(K\pi\pi\pi))$ 
is computed.
In the extraction procedure to determine $r_S$, we use the experimental distributions 
for the $r_D$ ratios, $\delta_D(K\pi)$, $k_D(K\pi\pi^0)$, $\delta_D(K\pi\pi^0)$, 
$k_D(K\pi\pi\pi)$ and $\delta_D(K\pi\pi\pi)$ 
\cite{ref:PDG,Asner:2008ft,ref:CLEOc}.  
All the remaining phases are extracted from a flat distribution in the range 
$[0,2\pi]$. 
$r_S$ is extracted from a flat distribution in the range 
$[0,1]$ and $k$ is extracted from a Gaussian distribution with mean 0.95 
and standard deviation 0.03.  
We obtain the likelihood ${\cal {L}}(r_S)$ shown in Fig.~\ref{fig:rS_threeCH}. 
The most probable value is $r_S=0.26$ and we obtain, by integrating the likelihood, 
the following 68$\%$ and 95$\%$ probability regions:
\begin{eqnarray}
&&r_{S} \in [0.18,0.34]\, @ \,68\% \mbox{ probability},\nonumber\\
&&r_S \in [0.07,0.41]\, @ \,95\% \mbox{ probability}. \nonumber
\label{eq:rSvalueALL}
\end{eqnarray}
Given the functional dependence of $R_{ADS}$ on $r_S$ ($R_{ADS}\sim r_S^2$), 
the likelihoods corresponding to $R_{ADS}<0$ have no effective role in the extraction of $r_S$. 
The dependence of the $r_S$ likelihood shown in Fig.~\ref{fig:rS_threeCH} on 
the choice of the prior distributions in the extraction procedure has been 
studied. 
While the 68$\%$ and 95$\%$ probability regions are quite stable, the 
likelihood shows a dependence on the choice of the prior distribution 
for values of $r_S$ close to zero.  
For this reason, the region near zero should not be used to 
evaluate the significance. The significance to observe $r_S$ different from 
zero corresponds to the significance for $R_{ADS}$, and is evaluated 
from the combined fit to be 2.5 standard deviations. 
The result obtained for $r_S$ with the procedure described above is consistent 
with the result found from a direct fit to data assuming the simplified 
expression $R_{ADS}=r_S^2$. 

In summary, we have presented a search for $b\to u$ transitions in 
$\Bz \rightarrow \Dtilde^0 \Kstarz $ decays, analysed through an ADS method. 
We see indications of a signal at the level of 2.5 standard deviations 
including systematic uncertainties. 
The most probable value for $r_S$ extracted from this result is $r_S=0.26$, 
where the 68$\%$ and 95$\%$ probability regions are indicated above. 
This result is in agreement with the phenomenological expectations from 
Ref. \cite{ref:cavoto}, and shows that the use of these decays and related ones
\cite{violaDalitz} for the determination of $\gamma$ is interesting in 
present and future facilities.


We are grateful for the 
extraordinary contributions of our \pep2\ colleagues in
achieving the excellent luminosity and machine conditions
that have made this work possible.
The success of this project also relies critically on the 
expertise and dedication of the computing organizations that 
support \babar.
The collaborating institutions wish to thank 
SLAC for its support and the kind hospitality extended to them. 
This work is supported by the
US Department of Energy
and National Science Foundation, the
Natural Sciences and Engineering Research Council (Canada),
the Commissariat \`a l'Energie Atomique and
Institut National de Physique Nucl\'eaire et de Physique des Particules
(France), the
Bundesministerium f\"ur Bildung und Forschung and
Deutsche Forschungsgemeinschaft
(Germany), the
Istituto Nazionale di Fisica Nucleare (Italy),
the Foundation for Fundamental Research on Matter (The Netherlands),
the Research Council of Norway, the
Ministry of Education and Science of the Russian Federation, 
Ministerio de Educaci\'on y Ciencia (Spain), and the
Science and Technology Facilities Council (United Kingdom).
Individuals have received support from 
the Marie-Curie IEF program (European Union) and
the A. P. Sloan Foundation.


\end{document}